\documentstyle[preprint,aps,epsf]{revtex}
\begin{document}
\tightenlines
\draft
\title{Bethe Approximation for  a Semi-flexible Polymer Chain}
\author{
Stefano Lise$^{1,2,a}$, Amos Maritan$^{1,2,b}$ and 
Alessandro Pelizzola$^{3,c}$ }
\address{(1) International School for Advanced Studies (SISSA),
 Via Beirut 2-4, 34014 Trieste, Italy  \\
and Istituto Nazionale per la Fisica della Materia (INFM)}
\address{(2) The Abdus Salam International Center for Theoretical
Physics, \\
Strada Costiera 11, 34100 Trieste, Italy}
\address{(3) Dipartimento di Fisica, Politecnico di Torino, 
c. Duca degli Abruzzi 24, 10129 Torino, Italy  \\
and Istituto Nazionale per la Fisica della Materia (INFM)}
\date{\today}
\maketitle
\begin{abstract}
We present a Bethe approximation to study lattice models of linear
polymers. The approach is variational in nature and based on the 
cluster variation method (CVM).
We focus on a model with $(i)$ a nearest neighbor attractive energy
$\epsilon _v$ between pair of non--bonded monomers, $(ii)$ a bending
energy $\epsilon _h$ for each pair  of successive chain segments 
which are not collinear.
We determine the phase diagram of the system as a function of the 
reduced temperature $t=\frac{T}{\epsilon _v}$ and of the parameter 
$x=\frac{\epsilon _h}{\epsilon _v}$.
We find two different qualitative behaviors, on varying $t$. For small
values of $x$ the system undergoes
a $\theta$ collapse from an extended coil to a compact globule;  
subsequently, on decreasing further $t$, there is a first order 
transition to an anisotropic phase, characterized by global orientational 
order. 
For sufficiently large values of $x$, instead, 
there is directly a first order transition from the coil to the orientational 
ordered phase. 
Our results are in good agreement with previous Monte Carlo simulations and
contradict in some aspects mean--field theory. 
In the limit of Hamiltonian walks, our approximation recovers results of
the Flory-Huggins theory for polymer melting.  
\vskip 0.6cm
\end{abstract}
\narrowtext
The configurational statistics of a long, linear polymer in solution
has often been modeled by a self-avoiding walk (SAW) on a lattice
\cite{des,vander}. The self--avoiding constraint takes into account
excluded--volume effects. Attractive Van der Waals interactions between
monomers are also generally considered. They are included by assigning a
negative energy $- \epsilon _v$ to each pair of nearest--neighbor monomers 
on the lattice, provided they are not consecutive along the chain. 
These interactions become relevant at low enough temperature, causing a 
collapse transition of the polymer\cite{deg1}. 
The transition point is called $\theta$ point and it 
separates a high temperature expanded structure from a low temperature 
compact globule.

Polymers with a local stiffness have also been considered, by including a 
bending energy $ \epsilon _h$ which favors straight segments  of the 
chain\cite{flory,kolinski,doniach,bastolla,doye}. 
The semi-flexible model has attracted much interest in the low temperature 
phase and  in particular in the limit of Hamiltonian walks, where  the 
path  is forced to visit all sites of  the lattice. In this case, it is 
believed to describe the melting of polymers chains\cite{flory}  
(see also \cite{nagle}). 
The system undergoes a phase transition between a disordered (liquid) phase 
and  an ordered (solid) phase, the latter being characterized by anisotropic
orientational order. 
More recently, the semi-flexible model has attracted renewed interest  
because of the possible  connection with the protein folding 
problem\cite{doniach,bastolla}. In this spirit, each link of the walk 
represents an $\alpha$--helical turn (ca. $3$ amino acids) and the curvature 
term mimics the tendency  to form  secondary structures. The attractive
energy between monomers models the hydrophobic effect, which is supposed
to be the main driving force for the folding transition\cite{dill}.


In the present work we consider a Bethe approximation for lattice
homopolymers. Our aim is twofold: first we introduce a new method to
deal with SAW problems, second we present results concerning the phase
diagram of an isolated semi-flexible chain.  In the approach to the
problem we have followed the cluster variation method 
(CVM)\cite{kik1,kik2,an}.
This is a closed form approximation, which is known to give excellent 
results for the phase diagram of spin systems\cite{cvm}.
The approximation scheme is determined by the largest
clusters of sites which are treated exactly. The CVM allows us to write
an approximate expression for the free energy of the system, as a function
of the probability of occurrence of all possible configurations of the
basic cluster. This free energy has then to be minimized, subject to 
consistency conditions on the distribution variables. 
The pair approximation considers nearest-neighbor pair of lattice sites and 
it corresponds to the Bethe approximation.


We represent a polymer as a SAW on a $d$--dimensional hyper-cubic lattice
with $V=L^d$ sites.
Pairs of non--consecutive vertices along the chain interact through an
attractive nearest--neighbor potential $\epsilon _v$. Stiffness of the chain 
is incorporated by attributing an energy penalty $\epsilon _h$ to each turn
(corner) of the walk. 
Let $T$ be the absolute temperature and $\beta=\frac{1}{\kappa _B T}$ the 
inverse temperature.
In the following we will adopt the notation $\omega= \beta \epsilon _v$,
$t=\frac{T}{\epsilon _v}$ and  $x=\frac{\epsilon _h}{\epsilon _v}$.
The partition function of the system is 
\begin{equation}
\label{partition}
Z_{\mbox{\tiny $N$}} = \sum _{\{SAW\}} e^{\omega ( N_{con}(\mbox{$\cal{S}$}) 
                          - x N_{cor}(\mbox{$\cal{S}$}))}
\end{equation}
where $\{SAW\}$ denotes the ensemble of all $N$--step SAW;
$N_{con}(\mbox{$\cal{S}$})$ and $N_{cor}(\mbox{$\cal{S}$})$
are respectively the number of contacts and corners in walk 
$\mbox{$\cal{S}$}$.
Introducing a monomer fugacity $z$, the grand
canonical partition function reads therefore
\begin{equation}
\label{granpart}
\mbox{$\cal{Z}$}=\sum _{N=1}^{\infty} z^N Z_{\mbox{\tiny $N$}}
\end{equation}
where the sum is over all possible lengths $N$ of the walk.
 

The lattice model (\ref{partition}) (or equivalently (\ref{granpart}))
has  been the  object of  recent investigations\cite{doniach,bastolla,doye}. 
A mean--field analysis\cite{doniach} predicts a
$\theta$--collapse transition at a temperature $t_{\mbox{\tiny $\theta$}}$ 
independent of $x$. Another transition should occur at 
lower temperature. It is a discontinuous melting transition from a disordered 
globule to an ordered ``crystalline'' phase. The melting temperature 
$t_{m}$ increases with $x$, although in mean-field approximation 
$t_{m} < t_{\mbox{\tiny $\theta$}}$ for any value of $x$.
This picture has been partly contradicted by heuristic 
arguments\cite{doniach,doye} and by Monte Carlo 
simulations\cite{bastolla,doye}. Indeed $t_{\mbox{\tiny $\theta$}}$ 
appears to slightly increase with $x$. 
Most importantly, $t_{m}$ seems to grow with stiffness
and it does not reach a finite asymptotic value. This implies that the 
line of the melting transition hits the line of the $\theta$--collapse. 
For sufficiently high values of $x$, there is therefore a direct first 
order transition from the open coil to the ordered phase. The triple
point is found to be approximately at $x \simeq 13$\cite{bastolla}.


The formulation of the CVM given by An~\cite{an} is a particularly
convenient starting point for our analysis. 
We first illustrate the method by treating in some detail the simpler 
case of zero stiffness ($x=0$). 
Let $s_i$ and $p_i$ be the distribution variables assigned respectively
to each site and pair configurations. One should in principle distinguish
among all the possible configurations which are not related by symmetry
operations. In $d=2$ there are, for instance, $3$ single site and $11$ pair
independent configurations. 
In fact we have verified that in the particular case $x=0$ they can be 
grouped  into a smaller number of non--equivalent classes. 
These are determined only by the following conditions: (a) the site is 
visited by the path, (b) the nearest--neighbor pair is joined directly by 
the path.
The independent configurations are reported schematically in fig.~\ref{fig1}, 
together with their multiplicity of  occurrence. 
Following the notation of fig.~\ref{fig1}, the free energy of the 
system (\ref{granpart}), in the pair approximation, can be written\cite{an}
\begin{eqnarray}
\label{free}
\frac{\beta F}{V} & = & - \frac{q(q-1)}{2} \ln z s_1 
-\frac{q(q-1)^2 (q-2)^2}{8} \omega p_2 + \nonumber \\
  & &   (1-q)\left( \sum _{i=1}^{2} m_s(i) s_i \ln s_i \right) + 
  \frac{q}{2}\left( \sum _{i=1}^{4} m_p(i) p_i \ln p_i \right) 
\end{eqnarray}
where $q=2 d$ is the coordination number of the lattice and $m_s(i)$ and
$ m_p(i)$ stand respectively for the multiplicity of site and pair  
configurations.
Normalization of the distributions and consistency conditions on the 
probability variables require respectively
\begin{equation}
\label{norm}
s_2  = 1 - \frac{q(q-1)}{2} s_1 
\end{equation}
and
\begin{eqnarray}
p_1 & = & \frac{s_1}{(q-1)} \nonumber  \\
\label{cons} 
p_2 & = & \frac{2 s_1}{(q-1)(q-2)} - \frac{2}{(q-1)(q-2)} p_3  \\  
p_4 & = & 1 - \frac{q(q-1)}{2} s_1 -\frac{(q-1)(q-2)}{2} p_3   \nonumber
\end{eqnarray}
This leaves us with only two variational parameters, e.g.  $s_1$ and $p_3$.
Substituting (\ref{norm}) and (\ref{cons}) into (\ref{free}) and 
minimizing with respect to $s_1$ and $p_3$ we obtain the stable phase 
at a given $z$ and $\omega$. 
We report the complete phase diagram for $d=3$ in fig.~\ref{fig2}. 
The polymer is a critical system along the transition line $z_c (\omega)$. 
This line separates a chain with zero density  
($s_1 =0$ for $z<z_c (\omega)$) from a chain with finite density  
($s_1 \ne 0$ for $z>z_c (\omega)$).
The continuous line represents a second order transition and the 
average number of monomers diverges with a power law as $z$ tends 
to $z_c$.
The broken line is instead a first order transition and  the density of 
monomers makes a finite jump at $z_c$. The cross denotes the tricritical 
point and it corresponds to the $\theta$ point\cite{deg2}.
In the case of pure SAW ($\omega=0$) the connectivity constant is 
$\mu = z_c^{-1}=2d-1$. 
This  result would have been expected by studying SAW on a Bethe lattice 
and it should be compared, for instance, with $\mu  \approx 2.64$\cite{conway} 
and $\mu  \approx 4.68$\cite{guttmann}, obtained from  exact enumerations 
respectively in $d=2$ and $d=3$. 
In our framework $z_c$ does not depend on $\omega$, as long 
as $\omega < \omega_{\mbox{\tiny $\theta$}}$. 
This is certainly an artifact of the approximation. It  can be ascribed 
to the fact that, after minimization in $p_3$, $p_2 \sim s_1^2$ in the limit
of $s_1$ going to zero. As a consequence, there is no term in (\ref{free}) 
proportional to $\omega$ which contributes in locating the minimum of the
free energy around $s_1=0$.
Nonetheless the estimates we obtain for the $\theta$ point  
$\omega _{\mbox{\tiny $\theta$}}^{\mbox{\tiny $(B)$}}$  are a better 
approximation to the available numerical values, with respect to mean 
field theory 
($ \omega_{\mbox{\tiny $\theta$}}^{\mbox{\tiny $(MF)$}}=
\frac{1}{2d}$\cite{doniach}):
in $d=2$, $\omega_{\mbox{\tiny $\theta$}} \approx 0.665$\cite{grass2}, 
$ \omega_{\mbox{\tiny $\theta$}}^{\mbox{\tiny $(B)$}} \approx 0.4055$ and  
$ \omega_{\mbox{\tiny $\theta$}}^{\mbox{\tiny $(MF)$}}=0.25 $;
in $d=3$, $ \omega_{\mbox{\tiny $\theta$}} \approx 0.275$\cite{tesi}, 
$ \omega_{\mbox{\tiny $\theta$}}^{\mbox{\tiny $(B)$}}\approx 0.2231$ and 
$ \omega_{\mbox{\tiny $\theta$}}^{\mbox{\tiny $(MF)$}}= 0.1667$.


In the general case $x \neq 0$ equations (\ref{free}), (\ref{norm}) and 
(\ref{cons}) must be generalized to include the curvature energy and 
the possibility of an anisotropic phase.
In this case it is not possible to group configurations as in fig.~\ref{fig1}
and one has to face a complex minimization problem.
An efficient way of doing it numerically is by mean of the natural iteration 
method\cite{kik2}. 
The resulting phase diagram is reported in fig.\ref{fig3}, 
as a function of $x$ and $t$ for $d=3$. The fugacity $z$ is fixed to its
critical value $z_c(x,t)$. This condition assures we are studying a polymer
in the limit of infinite chain length  ($N \rightarrow \infty$).
We find three different phases: an open coil, a compact globule and an 
ordered crystal. In our approximation, the latter is just the ground 
state of the polymer, having all links perfectly  aligned. 
This is known to be not completely correct, as it has been shown rigorously
that, for instance, in the case of Hamiltonian walks the entropy strictly 
vanishes only in the limit $T\rightarrow0$\cite{gujrati,goldstein}. 
The $\theta$--collapse line between the coil and the globule
appears to be independent of $x$. On the other hand the discontinuous 
melting transition tends to infinity with $x$. Beyond the triple point at
$x \simeq 8.8$, there is directly a first order transition from the coil to 
the solid. 


The limit $T \rightarrow 0$ (or $z \rightarrow \infty$) corresponds to
Hamiltonian walks. In this limit, walks are space filling and configurations 
with vacancies do not contribute to the partition function (\ref{granpart}).
Also, the attractive nearest--neighbor potential $\epsilon _v$ plays 
no role as there are precisely $(d-1)$ contacts per monomer. The resulting
model is the so called Flory model of polymer melting\cite{flory}. 
In this case, we obtain a minimum for the free energy in the disordered, 
compact phase which coincides with the free energy estimated by using the
Flory-Huggins approximation\cite{huggins}. 
The latter was originally derived from combinatorial arguments\cite{gibbs}. 
Neglecting the constant contribution proportional to $\epsilon _v$, the
analytical expression  reads
\begin{equation}
\label{FH}
\frac{\beta F_{\mbox{\tiny $FH$}} } {V}=
\ln \left[\frac{ \left(1-\frac{2}{q} \right)^{-\left(\frac{q}{2}-1 \right)}}
                {1+(q-2) \exp \left( - \omega x \right)} \right]       
\end{equation}
At low temperature $F_{\mbox{\tiny $FH$}}$  competes for stability with the 
local minimum associated to the ordered phase, which has strictly 
$F_{\mbox{\tiny $O$}} = 0$ in our
approach. A first order phase transition takes therefore place at 
\begin{equation}
t _m = \frac{x}{\ln \left[ 
\frac{q-2}{ \left(1-\frac{2}{q} \right)^{-\left(\frac{q}{2}-1 \right)}-1}
\right]}
\end{equation} 
In particular for $d=3$ ($q=6$) we have 
$t_m/x = \left(\ln\left( \frac{16}{5} \right)\right)^{-1} \approx  0.86 $,
which corresponds to the slope of the globule-solid transition line of 
fig.~\ref{fig3}, in the limit $x \rightarrow 0$.
This value is slightly larger than the analogous mean-field estimate,
 $t_m/x \approx 0.58$\cite{bascle}.

In the case $x=0$ each HW is equally weighted and the total number 
of paths is believed to scale as 
$\mbox{$\cal{N}$}_{\mbox{\tiny $HW$}}\simeq \mu_{\mbox{\tiny $H$}}^N$. 
From (\ref{FH}) we therefore obtain 
\begin{equation}
\label{muh_bet}
\mu_{\mbox{\tiny $H$}}^{\mbox{\tiny $(B)$}} = 
q \frac{q-1}{q-2} (1 - \frac{2}{q})^{\frac{q}{2}}
\end{equation}
A mean-field approach\cite{orland}, in very good agreement with numerical 
estimates, yields
\begin{equation} 
\label{muh_mf}
\mu _{\mbox{\tiny $H$}}^{\mbox{\tiny $(MF)$}} = \frac{q}{e}
\end{equation}
It is interesting to note that expression (\ref{muh_bet}) 
correctly  predicts $\mu _{\mbox{\tiny $H$}}=1$ for $d=1$, contrary to 
(\ref{muh_mf}).
In order to evaluate corrections to mean--field theory in powers of $1/q$, 
we have formulated  the Hamiltonian walk problem through spin variables.
This is in close analogy to the De Gennes theorem  for  SAW\cite{deg3}.
By using a suitable high--temperature expansion\cite{plefka} we have then 
been able to calculate the coefficients of the series up to the third order, 
extending of one order previous results\cite{nemirovsky}. We find
\begin{equation}
\label{mf_corr}
\mu_{\mbox{\tiny $H$}}=
\frac{q}{e}(1 + \frac{1}{6 q^2} - \frac{2}{ 3 q^3} + \ldots) 
\end{equation}  
In the limit $q \rightarrow \infty$ expression (\ref{muh_bet}) specializes to
\begin{equation}
\label{bet_corr}
\mu_{\mbox{\tiny $H$}}^{\mbox{\tiny $(B)$}} = 
\frac{q}{e}(1 + \frac{1}{6 q^2} + \frac{1}{3 q^3} + \ldots)
\end{equation}
The Bethe approximation therefore correctly recovers results of the expansion 
(\ref{mf_corr}) up to order $O(\frac{1}{q^2})$. 


In summary, we have presented a Bethe approximation for lattice
models of linear homopolymers. The method constitutes a substantial 
improvement with respect to mean--field theory. Indeed it produces a
phase diagram for a semi-flexible polymer chain which is in good 
agreement with Monte Carlo simulations. In particular, we find a 
triple point where the $\theta$ collapse line and the melting transition 
line meets.  
In the limit of Hamiltonian walks it recovers results of the Flory-Huggins 
theory for polymer melting, whose variational nature appears in a transparent 
shape within our framework.
It has the advantage of not requiring any spin or field theoretical 
representation, rather it relies directly on the configurations of the 
system. This last consideration suggests the scheme is more general and 
suitable to be applied to other geometrical problems, as, for instance, 
branched polymers\cite{delos} and self-avoiding surfaces\cite{gonnella}
(see \cite{vander} and references therein).  
It is plausible that the accuracy of the method  can be systematically 
refined according to the cluster variation method, in analogy with spin
systems. We expect some of the inaccurate features of the pair 
approximation could be removed by considering larger basic 
clusters.
\vskip 0.2cm


\noindent 
$^a$ Electronic address:       lise@sissa.it    \\
$^b$ Electronic address:       maritan@sissa.it \\
$^c$ Electronic address:       alex@athena.polito.it


\begin{figure}[hb]
\protect\vspace{0.3cm}
 \epsfxsize=3.3in
\centerline{\epsffile{fig1.eps}}
\protect\vspace{0.3cm}
\caption[1]{\label{fig1}
Schematic representation of independent (a) site
and (b) pair configurations, in the case of zero 
stiffness ($x=0$). The continuous line represents the path
visiting a site; $q=2d$ is the coordination number of the
lattice.}
\end{figure}

\newpage
\begin{figure}[hb]
\epsfxsize=3.7in
\centerline{ \epsffile{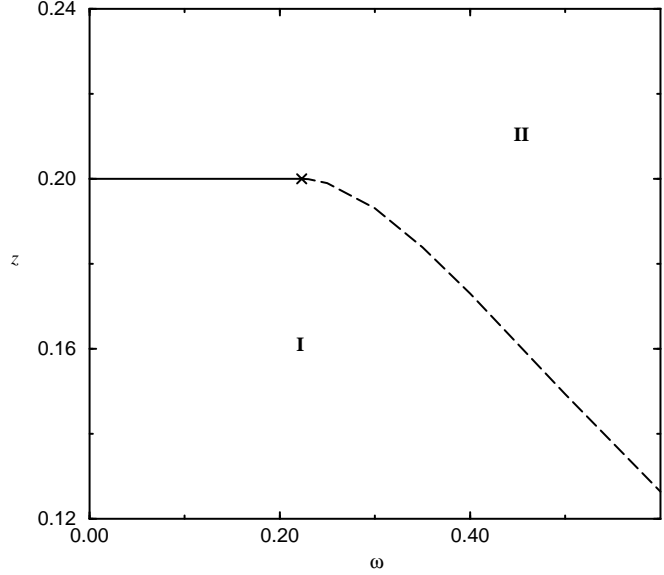}}
\caption[2]{\label{fig2} 
Phase diagram of the system as a function of $\omega$ and $z$,
in the case of zero stiffness ($x=0$). The average length of the
polymer is finite (infinite) in region I (II). 
The continuous (dashed) line is a second (first) 
order transition. The cross marks the tricritical point 
($\omega _{\protect\mbox{\protect\tiny $\theta$}}  \approx 0.2231$ and 
$z _{\protect\mbox{\protect\tiny $\theta$}} = 0.2$).}
\end{figure}

\begin{figure}[hb]
\epsfxsize=3.7in
\centerline{\epsffile{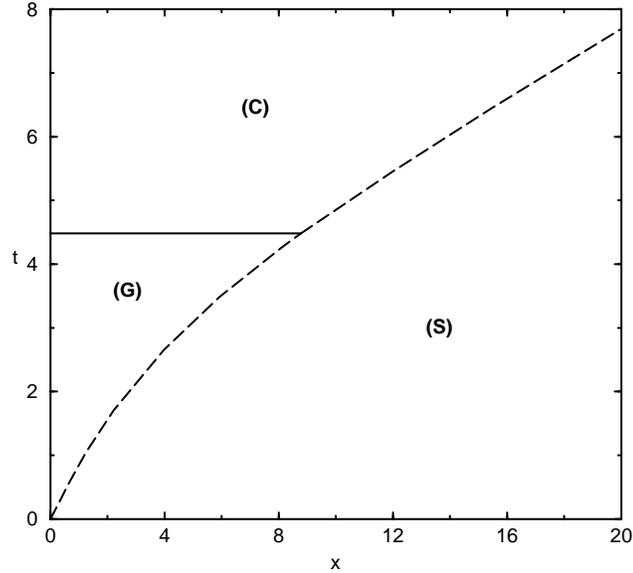}}
\caption[3]{\label{fig3}
Phase diagram of the system as a function of $x$ and $t$. 
The continuous line denotes the $\theta$ transition from
the coil (C) to the globule (G). The dashed line 
represents the first order transition to the solid (S).
The triple point is at $x \approx 8.8$ and $t \approx 4.5$.
See fig.~3 in \protect\cite{bastolla}  and fig.~8  in 
\protect\cite{doye}  for comparison with Monte Carlo 
simulations.}
\end{figure}

\end{document}